\documentclass{aa}  

\usepackage{graphicx}
\usepackage{textcomp}
\usepackage{multirow}
\usepackage{scrextend}
\usepackage[para]{threeparttable}

\usepackage{txfonts}
\begin{document}

   \title{Erosion of an exoplanetary atmosphere caused by stellar winds}

   \author{J. M. Rodr\'{\i}guez-Mozos
          \inst{1}
          \and
          A. Moya\inst{2,3}
          }
\authorrunning{Rodr\'{\i}guez-Mozos \& Moya}
   \institute{University of Granada (UGR). Dept. Theoretical Physics and Cosmology. 18071. Granada. Spain
   \and
   School of Physics and Astronomy, University of Birmingham, Edgbaston, Birmingham, B15 2TT, UK
    \and
    Stellar Astrophysics Centre, Department of Physics and Astronomy, Aarhus University, Ny Munkegade 120, DK-8000 Aarhus C, Denmark\\
             \email{A.Moya@bham.ac.uk}
             }

   \date{Received September 15, 1996; accepted March 16, 1997}

 
  \abstract
   {}
   {We present a formalism for a first-order estimation of the magnetosphere radius of exoplanets orbiting stars in the range from 0.08 to 1.3 M$_\odot$. With this radius, we estimate the atmospheric surface that is not protected from stellar winds. We have analyzed this unprotected surface for the most extreme environment for exoplanets: GKM-type and very low-mass stars at the two limits of the habitable zone. The estimated unprotected surface makes it possible to define a likelihood for an exoplanet to retain its atmosphere. This function can be incorporated into the new habitability index SEPHI.}
   {Using different formulations in the literature in addition to stellar and exoplanet physical characteristics, we estimated the stellar magnetic induction, the main characteristics of the stellar wind, and the different star-planet interaction regions (sub- and super-Alfv\'enic, sub- and supersonic). With this information, we can estimate the radius of the exoplanet magnetopause and thus the exoplanet unprotected surface.}
   {We have conducted a study of the auroral aperture angles for Earth-like exoplanets orbiting the habitable zone of its star, and found different behaviors depending on whether the star is in rotational saturated or unsaturated regimes, with angles of aperture of the auroral ring above or below 36$^\circ$, respectively, and with different slopes for the linear relation between the auroral aperture angle at the inner edge of the habitable zone versus the difference between auroral aperture angles at the two boundaries of the habitable zone. When the planet is tidally locked, the unprotected angle increases dramatically to values higher than 40$^\circ$ with a low likelihood of keeping its atmosphere. When the impact of stellar wind is produced in the sub-Alfvénic regime, the likelihood of keeping the atmosphere is almost zero for exoplanets orbiting very close to their star, regardless of whether they are saturated or not.}
   {}

   \keywords{Planets and satellites: magnetic fields --
                Planet-star interactions --
                Planets and satellites: atmospheres
               }

   \maketitle
%

\section{Introduction}

One of the characteristics that have received much attention in the current effort of searching for potentially habitable exoplanets is whether the planet is located in the habitable zone (HZ), allowing liquid water on its surface in first approximation \citep{Kopparapu14}. Additional physics in this direction, such as albedo or greenhouse effect, have also been taken into account. However, little attention has been paid in comparison to the effect of the space environment on the ability of the exoplanet to retain an atmosphere that is dense enough to protect any emerging life at its surface from stellar radiation.

Exoplanet atmospheres can be eroded by energetic radiation (UV and X-ray) and stellar winds \citep{Lammer03,Cohen15,Garraffo16}. This is particularly important for M-type and very low-mass (VLM) stars, where the HZ is located very close to the star, the stellar activity is usually significant, and the stellar wind is very dense.

When a planet is magnetized, the magnetosphere is defined as the part of space that is dominated by the planetary magnetic field, and the magnetopause as the outer boundary of the magnetosphere. The polar caps are the regions around the magnetic poles where the magnetic field lines are open; that is to say, where they have one end in the ionosphere and the other in interplanetary space \citep{Gunell18}. The magnetic field is a physical barrier that shields the exoplanet from this stellar environment, but the presence of a strong planetary magnetic field does not necessarily protect a planet from losing its atmosphere.

Thermal escape models suggest that exoplanetary atmospheres around active KM  stars undergo massive hydrogen escape. XUV radiation induces nonthermal heating through photoabsorption and photoionization that increase the temperature of the exosphere. At high XUV fluxes, this process initiates hydrodynamic atmospheric escape of neutral atmospheric species. Because it is the lightest component, hydrogen escapes faster than any other species through this mechanism \citep{Lammer08}

Other nonthermal models \citep{Airapetian17} suggest that the atmospheres of a significant fraction of Earth-like exoplanets orbiting M dwarfs and active K stars, which are exposed to high XUV fluxes, will have a significant atmospheric loss rate of oxygen and nitrogen, which will make them uninhabitable within a few tens to hundreds of megayears because of the low replenishment rate from volcanism or cometary bombardment.

On the other hand, it has been found that the atmospheric mass escape rate occurs in two areas. The first corresponds to polar cap escape and is dominant for hydrogen, while the second area is dominated by cusp escape \citep{Gunell18}. In the polar cap of a magnetized planet, magnetic field lines are opened, which allows the penetration of charged particles that are transported by the stellar wind, including cosmic rays, which can interact directly with the atmosphere of the planet. We call the fractional area of the planetary surface that has open magnetic field lines the unprotected surface.
When a star increases the magnetic and dynamic pressure on a planet, it not only produces a reduction of the radius of the magnetopause, but also an increase in unprotected surface of the planet \citep{Vidotto13}.

We here present a formalism based on current models for a first-order estimation of the magnetosphere radius of the exoplanet as a function of its orbit, its rotation period, its internal structure, and the stellar properties. With this radius, we can estimate the unprotected surface \citep{Vidotto13}. We have analyzed this unprotected surface for the most extreme environment for exoplanets: GKM-type and VLM stars at the two limits of the corresponding HZ. In addition, we have also studied some real cases.

Recently, \citet{SEPHI} have proposed a new statistically significant planetary habitability index where the exoplanet magnetic field plays a major role. The estimated unprotected surface together with some assumptions related to the ability of the Earth to retain its atmosphere with time makes it possible to define a likelihood function that can be incorporated into the index. This likelihood is related only to the atmosphere erosion caused by the stellar environment. It should be borne in mind that the thickness of the atmosphere has not yet been taken into account.

\section{Theoretical framework}

The study of the effect of stellar winds on exoplanetary atmospheres is the study of the interaction of three physical quantities. These are the stellar magnetic field and its translation into a stellar wind and radiation, the exoplanet magnetic field and its translation into a protective environment for the exoplanetary atmosphere, and the distance of the star to the exoplanet and its translation into a regime of sub-Alfv\'enic or super-Alfv\'enic, supersonic or subsonic stellar wind when it hits the magnetopause of the exoplanet.

\subsection{Alfv\'en radius}

The most important quantity for understanding the stellar wind velocity as a function of its distance from the star is the Alfv\'en radius. It is defined as the radial distance from the source star where the magnetic energy density is equal to the kinetic energy of the stellar wind \citep{Belenkaya}. Inside the Alfv\'en radius, the magnetic field transports the main part of the angular momentum of the stellar wind. Outside this radius, the angular momentum of the stellar wind is mainly due to the plasma movement.

Following \citet{Ardestani}, the Alfv\'en radius can be obtained as

\begin{equation}
    \frac{r_{\rm A}}{R_*}=\sqrt{\left|\frac{3\frac{dJ}{dt}}{2\dot{M}\Omega_*}\right|}
    \label{eq:alf_rad}
,\end{equation}

\noindent where $r_{\rm A}$ is the Alfv\'en radius, $R_*$ is the stellar radius, $\frac{dJ}{dt}$ is the stellar angular momentum loss with time, $\dot{M}$ is the stellar mass loss defined by equations \ref{Eq:mass_loss} and \ref{Eq:mass_loss_fx} depending on the stellar mass, and $\Omega_*$ is the stellar angular frequency.

These authors propose the following formulation for the angular momentum loss with time:

\begin{equation}
    \frac{dJ}{dt}=\frac{2}{3}K_1^2\dot{M}^{(1-2m)}R_*^{(2+4m)}B_0^{4m}
    \frac{\Omega_*}{\left( K_2^2\nu_{\rm esc}^2+\Omega_*^2R_*^2 \right)^m}
,\end{equation}

\noindent where $K_1$, $K_2$, and $m$ are unknown constants, $B_0$ is the stellar magnetic induction, $\nu_{\rm esc}=\sqrt{{\rm G}M_*/R_*}$ is the stellar escape velocity, $M_*$ is the stellar mass, and G is the gravitational constant. One of the main assumptions for deriving this expression is that the stellar wind is isotropic, that is,

\begin{equation}
    \dot{M}=4\pi\,r^2\,\rho(r)\,u(r)
\label{eq:isot}
,\end{equation}

\noindent where $r$ is the distance from the stellar center, $\rho(r)$ is the stellar wind density, and $u(r)$ the stellar wind velocity.

Although we can find some alternative expressions for the stellar rotation breaking \citep[see, e.g.,][]{vanSaders13,Matt12,Matt15,Garraffo18}, we have decided to follow the expression in \citet{Ardestani} in coherence with the rest of our work. The results shown in Section 4 validate this assumption. Sadeghi\ Ardestani and collaborators have empirically calibrated the unknown constants and obtained $K_1=6.43$, $K_2=0.0506$, and $m=0.2177$. \citet{Matt12} proposed a similar angular frequency evolution law, and they also empirically calibrated their unknown constants. In both cases, $K_2$ and $m$ account for the same functional dependence, and both teams reach the same empirical values for these two constants.

For $\dot{M}$ and $B_{\rm 0}$, \citet{Pallavicini} found a relation between the stellar X-ray emission relative to the bolometric luminosity ($L_{\rm X}/L_{\rm bol}$) and its rotation. This relation was the origin of the relation of stellar activity to rotation and its connection with stellar dating, which has been widely studied since then. For rapid rotators, that is, stars with a Rossby number lower than 0.1, this relation breaks and remains constant around $L_{\rm X}/L_{\rm bol}\approx 10^{-3}$ \citep{Vilhu, Micela, Wright18}. This is the so-called saturated regime. On the other hand, slow rotators are in the unsaturated regime where $L_{\rm X}/L_{\rm bol}$ decrease with the stellar rotation \citep{Reiners}.




M dwarfs and VLM stars, as any other stars, can be fast or slow rotators. From a statistical point of view, however, they are fast rotators when they are young and eventually reach the expected rotational velocity predicted by the breaking laws \citep{Matt15, Garraffo18}. As the star evolves, it tends to lose angular momentum with time, which breaks its angular velocity down. This shows that a star can change its regime from saturated to unsaturated during its life. This transition occurs when the Rossby number ($R_{\rm o}$) has a value of about 0.1 \citep{Wright11} and depends on different factors such as the extension of the stellar convective zone. In this work, the stellar angular frequency at which this transition occurs is estimated following the empirical relation found by \citet{Johnstone_Gudel},

\begin{equation}
    \Omega_{\rm sat} = 13.53\,\Omega_\odot\left(\frac{M_*}{M_\odot}\right)^{1.08}
    \label{eq:saturation}
,\end{equation}

\noindent where $\Omega_{\rm sat}$ is the minimum angular velocity that defines the saturated regime, $M_\odot$ represents the solar mass, and $\Omega_\odot$ is the solar rotational angular frequency.

\citet{Ardestani} proposed a functional form for estimating $\dot{M}$ and $B_{\rm 0}$ from solar values depending on the stellar Rossby number and its stellar activity - rotation regime as

\begin{eqnarray}
     \dot{M} = \dot{M}_\odot \left(\frac{R_{{\rm o},\odot}}{R_{\rm a}}\right)^{1.3} \label{Eq:mass_loss}\\
     B_{\rm 0} = B_\odot \left(\frac{R_{{\rm o},\odot}}{R_{\rm a}}\right)^{1.2}
     \label{Eq:Momento}
,\end{eqnarray}

\noindent where $R_{{\rm o},\odot}$, $B_\odot$, and $\dot{M}_\odot$ are the solar Rossby number, magnetic induction, and mass loss, respectively,

\begin{equation}
    R_{\rm a} = 
\begin{cases}
    R_{\rm o*},& \text{if } \Omega_* \leq \Omega_{\rm sat}\\
    0.09,& \text{if } \Omega_* > \Omega_{\rm sat}
\end{cases}
\end{equation}

\noindent and

\begin{equation}
    R_{o*}=\frac{2\pi}{\Omega_*\tau_{\rm c}}
    \label{eq:rossby}
\end{equation}

\noindent is the stellar Rossby number, with $\tau_{\rm c}$ the convective turnover time. The estimation of $\tau_{\rm c}$ has been addressed by different authors \citep{Wright11, Wright18}. In order to be coherent, we use the dependence of $\tau_{\rm c}$ on stellar mass ($M_*$) and radius ($R_*$) presented by \citet{Ardestani},

\begin{equation}
    \tau_{\rm c} \propto M_*^{-1}R_*^{-1.2}
    \label{eq:tau}
.\end{equation}

In that study, the authors develop a grid of angular momentum evolution models for stars with masses in the range [0.5, 1.6] ${\rm M}_\odot$ whose fundamental variable is the Rossby number. Related to its top limit, in this work we study stars with a developed outer convective zone, that is, with masses below 1.3M$_\odot$. On the other hand, the bottom limit of 0.5M$_\odot$ is only a consequence of the grid of models used by \citet{Ardestani}. Their Figure A2 shows that Equation \ref{eq:tau} could be extended to lower masses. In Section 4.2 we present a comparison between the predicted $R_{\rm o}$ using this equation and observed Rossby number for a set of stars with masses in the range of 0.08 to 1 M$_\odot$.

However, the estimation of the mass loss as a function of the Rossby number of \citet{Ardestani} (Eq. \ref{Eq:mass_loss}) is only valid for masses in the range from 0.5 to 1.6 $M_\odot$. For lower stellar masses, \citet{Wood02,Wood05} found a relation between the X-ray flux and the mass loss

\begin{equation}
    F_X^{1.34} \propto \frac{\dot{M}}{4\pi R_*^2}
    \label{Eq:mass_loss_fx}
.\end{equation}

This expression is valid for low-mass stars older than 600 Myr and $F_X<10^6$ ergs cm$^{-2}$ s$^{-1}$. This means that the younger the star, the higher the mass loss, in coherence with the bases of gyrochronology and the activity - age relation.

To evaluate the effect of using this formalism to estimate the Alfv\'en radius, we compared its predictions for the Sun, Proxima Cen., and TRAPPIST-1 (30, 25, and 63 $R_*,$ respectively) with 24 R$_\odot$ for the Sun and the results of \citet{Garraffo16, Garraffo17} using 3D models for Proxima Cen. (33R$_*$ with a B$_o$=600 G), and for TRAPPIST-1 (53 R$_*$ for B$_o$=300G and 75 R$_*$ for B$_o$=600G). The former estimates are comparable to the latter estimates.

\subsection{Stellar wind velocity and sound radius}

The velocity at which the stellar wind impacts the planet is a critical value for estimating the erosion that it can produce in the planet atmosphere. Using spherical coordinates according to the assumption of an isotropic stellar wind, this velocity can be expressed as a radial plus an azimuthal component,

\begin{equation}
    u(\vec{r}) = u_r\,\hat{r} + u_\theta \hat{\theta}
.\end{equation}

The estimation of these components depends on the Alfv\'en radius and the sound velocity. The interaction of the stellar wind with the planetary magnetosphere produces bow shocks in super-Alfv\'enic conditions, but in sub-Alfv\'enic conditions, these interactions produce Alfv\'en waves \citep{Vidotto13}. On the other hand, \citet{Cohen14} found that the absence of bow shocks in the subsonic stellar wind is translated into a much deeper penetration of this wind in the planetary magnetosphere.

The sound velocity of the stellar wind $u_{\rm o}$, following the Parker's isotherm model \citep{Parker58}, depends on the stellar coronal temperature $T_{\rm o}$, and their solar values $u_{{\rm o}, \odot}$, and $T_{{\rm o}, \odot}$, as

\begin{equation}
    u_{\rm o} = u_{{\rm o}, \odot}\sqrt{\frac{T_{\rm o}}{T_{{\rm o},\odot}}}
.\end{equation}

\citet{Johnstone_Gudel} found the following relation between $T_{\rm o}$ and the X-ray flux:

\begin{equation}
    T_{\rm o} = 0.11\,F_{\rm x}^{0.26}
,\end{equation}

\noindent and then, using the results of \citet{Reiners}, these authors found that the coronal temperature can be expressed as

\begin{equation}
    T_{\rm o} = T_{{\rm o}, \odot}\,M_*^{-0.42}\left(\frac{\Omega^\prime}{\Omega_\odot}\right)^{0.52}
,\end{equation}

\noindent where

\begin{equation}
    \Omega^\prime = 
\begin{cases}
    \Omega_*,& \text{if } \Omega_* \leq \Omega_{\rm sat}\\
    \Omega_{\rm sat},& \text{if } \Omega_* > \Omega_{\rm sat}
\end{cases}
.\end{equation}

Moreover, the radial distance where the stellar wind changes from the subsonic regime to supersonic can be expressed, following that study, as

\begin{eqnarray}
     r_{\rm s} = r_{{\rm s},\odot}\,M_*\left(\frac{T_{{\rm o},\odot}}{T_{\rm o}}\right)
,\end{eqnarray}

\noindent where $r_{{\rm s}, \odot}$ is the corresponding solar value.

Following \citet{Parker58}, the radial component of the stellar wind can be expressed as

\begin{eqnarray}
     u_{\rm r}=2\,u_{\rm o}\sqrt{\ln{\frac{r}{r_{\rm s}}}}
\end{eqnarray}

\noindent for $r>>r_s$, where $r$ is the distance to the star, and

\begin{eqnarray}
     u_{\rm r}=e^{3/2}\,u_o\,e^{-\frac{2\,r_s}{r}}
\end{eqnarray}

\noindent for the rest of the radii.

In our case, the subsonic regime is usually located very close to the star. Therefore, all the exoplanets we study orbit in the supersonic regime.

To estimate the azimuthal component, we must take into account that the stellar wind follows the magnetic field lines. It will depart from the stellar surface with initial null velocity and it will follow the open magnetic field lines pushed by the magnetic pressure. Within the Alfv\'en radius, the azimuthal component of the stellar wind is proportional to the distance to the star, reaching its maximum at the Alfv\'en radius itself:

\begin{equation}
    u_\theta = \Omega_* r
.\end{equation}

From this radius on, this azimuthal component decreases following the inverse of the distance.

\subsection{Planetary magnetic induction}

Planetary magnetic induction cannot be obtained directly, except for the case of planets in the solar system. For this estimation, we used the theoretical works of \citet{Olson06}, \citet{Zuluaga13}, and \citet{Zuluaga18}. In the dipolar regime, the planetary magnetic induction can be estimated using the results of \citet{Olson06} as

\begin{equation}
    B_{\rm dip} = B_\oplus\left(\frac{\mathcal{M}_{\rm p}}{\mathcal{M}_\oplus}\right)\left(\frac{r_{\rm o}}{r_{{\rm o}, \oplus}}\right)^{-3}
    \label{campo_mag_polos}
,\end{equation}

\noindent where $r_{\rm o}$ is the core radius of the exoplanet, $\mathcal{M}_\text{p}$ is the magnetic moment of the exoplanet, and $B_{\oplus}$, $\mathcal{M}_\oplus$, and $r_{{\rm o}, \oplus}$ are the dipolar magnetic induction, dipolar magnetic moment, and core radius of Earth, respectively.

The core radius of the exoplanet can be estimated for dry telluric exoplanets with known mass and radius using the preliminary reference Earth model \citep[PREM,][]{Zeng16}. On the other hand, the exoplanet magnetic moment can be also estimated following \citet{Olson06} depending on whether the planet is tidally locked to the star, in which case we know its rotational velocity, or if it rotates freely, in which case its magnetic moment is independent of this rotational velocity.

\subsection{Planetary magnetopause radius}

The magnetosphere acts as a shield that redirects the particles of the stellar wind, prevents the direct interaction of this wind with the planetary atmosphere and  thus its erosion \citep{Vidotto18}. The stellar magnetic pressure and/or the dynamical pressure of the stellar wind can reduce the planetary magnetopause size, which exposes a significant fraction of the planetary atmosphere to erosion by the stellar wind \citep{Vidotto13}. When the planetary magnetopause is close to the planetary radius, the solar particles directly hit the planetary atmosphere. This can occur in particular when the planet is exposed to extreme wind pressure in sub-Alfv\'enic regions \citep{Garraffo17}.

To obtain the radius of the planetary magnetopause, we can use the expression of \citet{Vidotto18},

\begin{equation}
    \frac{r_{\rm M}}{R_{\rm P}} = \left[\frac{B^2_{\rm p,eq}}{8\pi\,P_{\rm dyn}+B^2_{\rm sa}}\right]^\frac{1}{6}
    \label{plan_mag_rad}
.\end{equation}

This equation is valid for planets that are magnetized with a dipolar magnetic field parallel to the rotation axis of the star, where the thermal pressure has been neglected, $r_{\rm M}$ is the radius of the planetary magnetopause, $R_{\rm P}$ is the planetary radius, $B_{\rm p,eq}$ is the planetary magnetic induction at the equator, $P_{\rm dyn}$ is the dynamical pressure of the stellar wind, and $B_{\rm sa}$ is the stellar magnetic induction at the planetary orbit.

$P_{\rm dyn}$ can be defined as

\begin{equation}
    P_{\rm dyn} \sim \rho (a)[\Delta u(a)]^2
,\end{equation}

\noindent where $a$ is the orbital distance, $\rho (a)$ can be obtained from the assumption of an isotropic stellar wind (Eq. \ref{eq:isot}), and $\Delta u$ accounts for the velocity of the planet relative to the stellar wind. For planets in close-in orbits, the high orbital velocity can cause azimuthal supersonic impacts even though the stellar wind velocity is low and the wind is still being accelerated \citep{Vidotto10a}. When the predominant impact occurs in the azimuthal direction and the orbital spin is coincident with the stellar rotation spin,

\begin{equation}
    \Delta u(a)=u_\theta (a) + v_{\rm t}
,\end{equation}

\noindent where $v_{\rm t}$ is the translation velocity of the planet. On the other hand, for radial impacts,

\begin{equation}
    \Delta u(a) = u_{\rm r} (a)
.\end{equation}

Close-in planets orbiting near the Alfv\'en radius of a fast-rotating star for example are therefore in a scenario of high mass loss, high azimuthal velocity, and high orbital velocity. All this causes a dense stellar wind that generates a high dynamical pressure on the planetary atmosphere.

To estimate the magnetic induction of the planet in Eq. \ref{plan_mag_rad}, we use the expressions shown in \citet{Vidotto13},

\begin{equation}
     B_{\rm p,eq} \approx \frac{1}{2}\,B_{\rm pp}
,\end{equation}

\noindent where $B_{\rm pp}$ is the planetary magnetic induction at the planet poles. The radial magnetic induction produced by the star on the orbit of the planet can be determined using the following expression:

\begin{equation}
     B_{\rm sa} = B_{\rm o}\left(\frac{r_{\rm o}}{a}\right)^2
,\end{equation}

\noindent where $r_{\rm o}$ is the radius of the stellar corona and $B_{\rm o}$ is the radial component of the magnetic field at the base of the corona.

The geometrical form of the magnetic field means that the magnetic poles of the planet are usually unprotected, and the stellar wind and cosmic radiation can directly interact with the planetary atmosphere. This unprotected zone has at first order the form of a cone of revolution whose axis is the magnetic axis of the planet and angle $\alpha_0$. \citet{Vidotto13} described this angle as a function of the planetary magnetopause radius as

\begin{equation}
    \alpha_0 = {\rm arc\,sin}\left[\left(\frac{R_{\rm p}}{r_{\rm M}}\right)^\frac{1}{2}\right]
.\end{equation}

For example, the Earth, with an $r_{\rm M}\approx 10R_{\oplus}$ \citep{Kivelson07}, currently has an unprotected angle $\alpha_0=18^\circ$. If the magnetic moment of the planet is too weak compared to the dynamic and magnetic pressures, the resulting situation is equivalent to the planet being demagnetized. A demagnetized Earth-like planet orbiting an M-type star can lose all its ozone in a very short time, allowing high-energy cosmic particles to reach the planetary surface in great numbers \citep{Tilley19}.

\section{Likelihood of a planet to retain its atmosphere as a consequence of the stellar wind}

It is currently not possible to determine exactly whether a planet is able to retain its atmosphere under a certain environment. The number of assumptions and uncertainties dominates the study. We can estimate the likelihood of the planet to retain its atmosphere, however, for which we used current standard observations and the formulation we presented here. An additional benefit of estimating this likelihood is that it can be included in the new statistical likelihood of exoplanetary habitability index, SEPHI \citep{SEPHI}.

We do not know the maximum unprotected atmosphere surface that allows a magnetized planet under significant pressure from the stellar wind to keep its atmosphere. Our only reference is Earth. \citet{Tarduno10} studied the "young Earth" case when our planet was around 1Gyr old. They estimated that at this time, the magnetosphere radius of Earth was approximately $r_{\rm M}\approx 5R_\oplus$ during millions or tens of million years, that is,  $\alpha_0 \approx 27^\circ$. Even in this scenario, the atmosphere loss of Earth was not enough to evaporate it completely.

Similarly, we do not know whether this young Earth scenario would finally end in complete atmosphere evaporation if this condition was maintained for a longer time. To allow a likelihood of 1 for a planet to retain its atmosphere due to space weather, we therefore adopted a conservative unprotected angle of $\alpha_{0,{\rm young\,earth}} \approx 25^\circ$, which corresponds to a minimum planetary magnetosphere radius $[r_{\rm M \oplus}/R_\oplus]_{\rm young\,earth} \sim 5.4$, that is, a total unprotected surface of 9$\%$.

On the other hand, \citet{Lammer07a} proposed that the radius of the planetary magnetosphere must be at least $[r_{\rm M}/R_{\rm P}]_{\rm min} =2$ for an efficient protection of its atmosphere, that is, $\alpha_{0, {\rm max}} \approx 45^\circ$. Nevertheless, for highly magnetized super-Earth planets, the high gravity works against the thermal escape of the lighter elements. We therefore chose a more conservative criterion for a null likelihood to retain an atmosphere due to stellar winds: $\alpha_{0, \rm max} \approx 50^\circ$, or $[r_{\rm M}/R_{\rm P}]_{\rm max} \sim 1.7$, that is, 36$\%$ of the surface is not protected.

With all this, we can define the total likelihood of a planet to retain its atmosphere due to space weather, compared with Earth, as

\begin{equation}
    P(x) = 
\begin{cases}
    1,& \text{if } \alpha_{\rm 0} \leq \alpha_{0, \rm young\, earth}\\
    e^{-\frac{1}{2}\left[\frac{\alpha_{\rm 0}-\alpha_{0, \rm young\, earth}}{\sigma}\right]^2},& \text{if } \alpha_{\rm 0} > \alpha_{0, \rm young\, earth}
\end{cases}
\end{equation}

\noindent with $\sigma=\frac{\alpha_{0, \rm max}-\alpha_{0, \rm young\, earth}}{3}$.

We recall that to obtain this probability, the following assumptions have been made:

\begin{itemize}
    \item The planet is magnetized with a dipolar magnetic field whose magnetic induction at the poles and semimajor axis are known.
    \item The star has an outer convective layer, that is, a stellar mass of between 0.08 and 1.3 M$_\odot$ , and its rotation axis is aligned with the magnetic axis of the planet. We also know the stellar mass, radius, and rotational period.
    \item The stellar wind is isotropic and isothermal.
\end{itemize}

If the magnetic axis of the planet is not aligned with the rotational axis of the star, forming an angle $\iota$, to calculate the radius of the magnetopause, we need to consider that the component of the magnetic pressure generated by the planet in the direction of the equatorial plane of the star is reduced by a factor cos $\iota$. In this way, the magnetopause values obtained using equation \ref{plan_mag_rad} are significantly affected by this correction in the case of a large misalignment (larger that 30$^\circ$ approximately). Unfortunately, current exoplanet detection techniques do not allow us to estimate $\iota$. We therefore used a general value of $\iota = 0^\circ$ as the most likely value for planets orbiting stars within the mass range of our study.

\section{Discussion}

To analyze the physical information we can extract from our theoretical framework, we studied different planetary system structures with different physical characteristics. The solar values we used for the different variables are shown in Table \ref{Tab:solar}.

\begin{table}
	\centering
	\caption{Solar values. \label{Tab:solar}
	}
	\begin{tabular}{ccc}
		\hline
		Variable  & Value & Unit \\
		\hline
		P$_\odot$ & 28 & days \\
		$\Omega_\odot$ & 2.6 $10^{-6}$ & Hz\\
		$\tau_{{\rm c}\,\odot}$ & 12.9 & days \\
		R$_{{\rm o}\,\odot}$ & 2.17 \\
		$\dot{M}_\odot$ & 2 $10^{-14}$ & M$_\odot$ yr$^{-1}$ \\
		T$_{{\rm o}\,\odot}$ & 2 $10^{6}$ & K \\
		B$_{{\rm o}\,\odot}$ & 1 & G\\
		u$_{{\rm o}\,\odot}$ & 180 & Km s$^{-1}$\\
		r$_{{\rm s}\,\odot}$ & 2 $10^{6}$ & Km \\
		\hline
	\end{tabular}
\end{table}

\subsection{Application to solar system planets}

We tested the accuracy of the formalism we presented in Section 2 with the magnetized planets of the solar system. The value of the angle $\iota$ was considered in this case because it is known. The results are summarized in Table \ref{Tab:solar_planets}. This table shows that the concordance between our predictions for the magnetopause radius and the observed values is remarkable.

\begin{table}
	\centering
	\caption{Radius of the magnetopuase for planets of the solar system. $B_{\rm pp}$ is the polar magentic induction of a planet in Gauss. i is the angle between the solar rotational axis and the magnetic axis of the planet. $r_{\rm M}/R_{\rm P}$ is the radius of the planetary magnetopause relative to the planetary radius as obtained with our formalism. $r_{\rm M}/R_{\rm P}$ (obs) is the observed radius of the planetary magnetopause relative to the planetary radius \citep{Kivelson07}. $\alpha_{\rm 0, theor}$ is the aperture angle of the auroral ring as obtained with our $r_{\rm M}/R_{\rm P}$. \label{Tab:solar_planets}
	}
	\begin{tabular}{cccccc}
		\hline
		\multirow{2}{*}{Planet}  & $B_{\rm pp}$ & $\iota$ &  $r_{\rm M}/R_{\rm P}$ & $r_{\rm M}/R_{\rm P}$ & $\alpha_{\rm 0, theor}$\\
		& (G) & ($^\circ$) & & (obs) & ($^\circ$)\\
		\hline
		Mercury & 0.007 & 0 & 1.4 & 1.4 & 59\\
		Earth & 1 & 12 & 10 & 8-12 & 18\\
		Jupiter & 8.56 & 10 & 51 & 50-100 & 8\\
		Saturn & 0.42 & 27 & 21 & 16-22 & 12\\
		Uranus & 0.23 & 0-90 & 21 & 18 & 12\\
		Neptune & 0.28 & 0-76 & 26 & 23-26 & 11\\
		\hline
	\end{tabular}
\end{table}

For ice giant planets, the magnetic dipole is not aligned in general with the planet spin axis. For Uranus it is displaced by 59$^\circ$ , and its center is markedly displaced from the planet center by about 0.3 planetary radii. The situation is similar for Neptune, with a displacement of 47$^\circ$ and 0.55 planetary radii. This means that $\iota$ does not only depend on the obliquity of the planet and the tilt angle of the dipole moment with respect to the planetary spin axis, but also on the phase angle of the planet in orbit and the phase angle of the planetary spin. For these planets, $\iota$ is therefore a function of time \citep{Lepping94}.

\subsection{Atmospheric erosion of Earth-like planets in the HZ}
\label{sec:ELHZ}

Our second analysis was made using the set of 33 testing stars whose main physical characteristics are summarized in Table \ref{Tab:earth_like_data}. This set was formed from the stars studied by \citet{Vidotto13}, M-type stars with planets, almost all in saturated regime, other stars with confirmed telluric planets, and the Sun as the reference. \citet{Vidotto13} did not provide the uncertainties for these stars, therefore we chose not to include them in the analysis of this section. To show the effect of the observational errors, a proper uncertainty analysis is included in the studies of Section 4.4.

When we compare the Rossby number obtained by applying equation \ref{eq:tau} with the values available in the literature, the mean error is 0.006$\pm$0.012. This validates the use of the formalism described at Section 2.1 and the assumptions therein. On the other hand, to test the accuracy of equation \ref{Eq:Momento}, we estimated the magnetopause radius obtained using the observed magnetic field in Table \ref{Tab:earth_like_data} for the set of \citet{Vidotto13} alone and for the set obtained using this equation. The mean difference found at the inner boundary of the HZ, where the effect is strongest, is 0.19 $r_{\rm M}/R_{\rm P}$. This uncertainty translates into an uncertainty in the determination of the angle of the auroral ring of 0.35$^\circ$ and a probability uncertainty of 0.05$\pm$0.06. These two uncertainties are negligible compared with the differences we found that are caused by the different physical environments.

To understand the effect of these stellar winds on potential Earth-like planets at the HZ, we assumed an Earth in terms of mass ($M_{\rm P}=1M_\oplus$), radius ($R_{\rm P}=1R_\oplus$), and polar magnetic field ($B_{\rm po} = 1G$) orbiting at the two HZ boundaries (D$_2$, inner HZ boundary; D$_3$, outer HZ boundary). D$_2$ and D$_3$ were estimated using the equations published by \citet{Kopparapu13, Kopparapu14}.

The results obtained at D$_2$ are summarized in Table \ref{Tab:earth_like}. In all cases, the impact type of the stellar wind is superssonic and super-Alfv\'enic, except for WX Uma, DX Cnc, and Trappist-1, where the regime is sub-Alfv\'enic. For all the stars in saturated regime, the unprotected angle is larger than 36$^\circ$, with a probability for retaining an atmosphere lower than 40$\%$. Stars that are more massive and/or slow rotating have a lower impact on the exoplanet atmosphere.

In Table \ref{Tab:earth_like} we also show the results obtained at D$_3$. The impact type of the wind is supersonic and super-Alfv\'enic in all cases. The protected surface is larger than that at D$_2$, as expected. The probability for retaining an atmosphere is higher than 50$\%,$ except for the lowest mass stars. In all analyzed cases, and at D$_2$ and D$_3$, the dynamic pressure produced by the stellar wind on the planet is a fundamental factor that cannot be neglected.

\begin{figure*}
 \includegraphics[width=\linewidth]{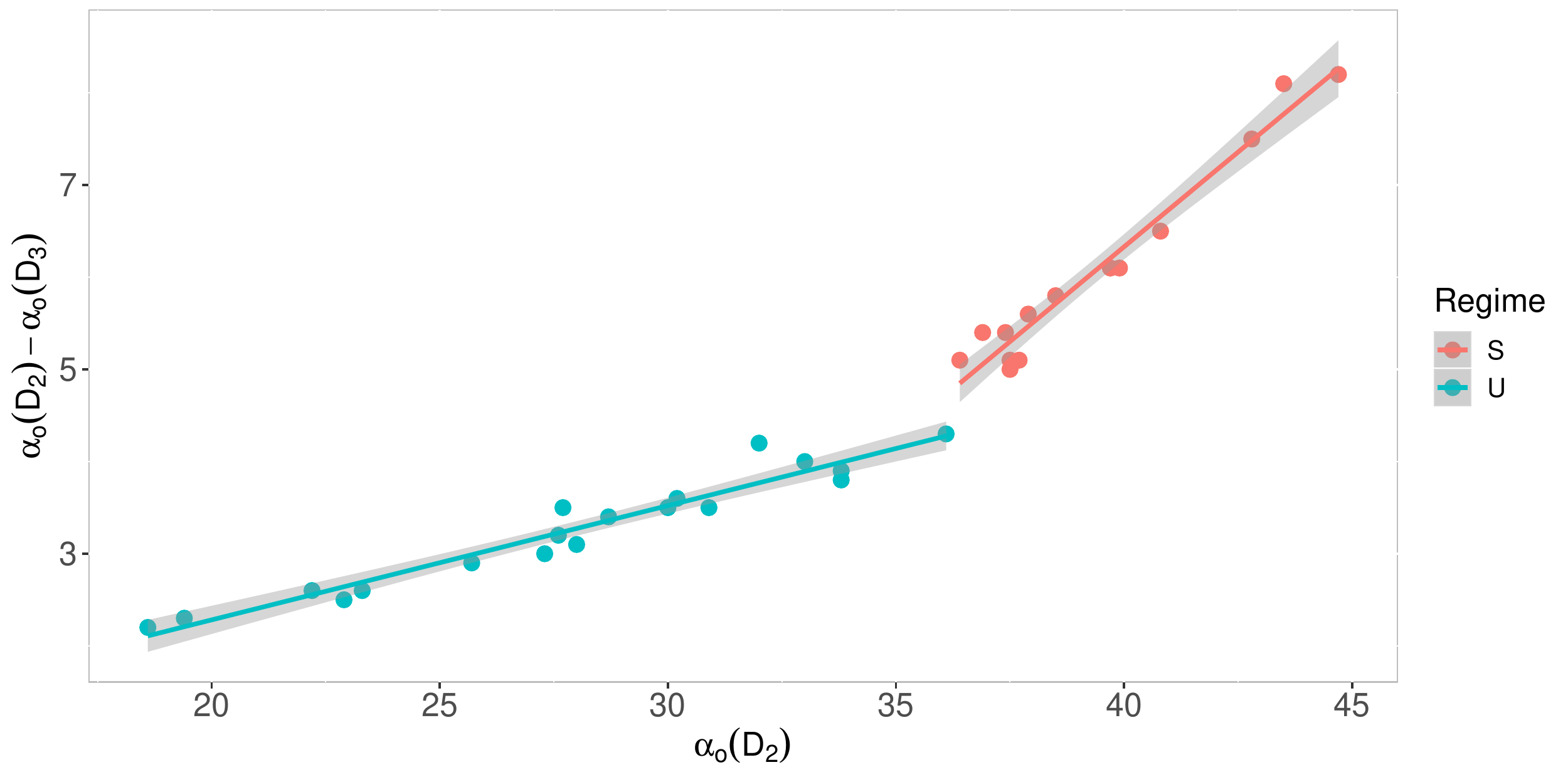}
 \caption{$\alpha_{\rm 0}$ at D$_2$ vs. $\alpha_{\rm 0}$ difference between the two HZ boundaries.}
 \label{Fig:alpha_evol}
\end{figure*}

In Fig. \ref{Fig:alpha_evol} we show the difference between the auroral rings at D$_3$ and D$_2$ as a function of the auroral ring at D$_2$. Two different behaviors were found. On the one hand, planets orbiting saturated stars (S, red) have a larger unprotected zone ($\alpha_{\rm o} > 36^\circ$) than those orbiting unsaturated stars (U, blue, $\alpha_{\rm o} < 36^\circ$) regardless of the mass of the star. On the another hand, there appears to be a linear relationship between $\alpha_{\rm o}({\rm D}_2)$ and $\alpha_{\rm o}({\rm D}_2)$ - $\alpha_{\rm o}({\rm D}_3)$ with different slopes depending on the stellar regime.

\subsection{Effect of tidal-locking in planet shielding}

The analysis in the previous section was made assuming an Earth-like planet rotating freely at the inner and outer edges of the HZ. However, for low-mass red dwarf stars, the HZ is very close to it and the exoplanets in this area are in many cases tidally locked. In general, tidally locked exoplanets may rotate more slowly and their own magnetic field and the unprotected zone of their atmosphere can therefore be affected.

In this section we analyze the effect of this tidal locking on planetary atmospheres. Whether a planet is tidally locked to its star depends mainly on its orbital distance and the stellar mass. Therefore, and because of the lack of observational data that properly cover the entire stellar mass space in this case, we simulated a set of stars, both saturated and unsaturated, with masses ranging between 0.08 and 1 M$_\odot$ in steps of 0.1 M$_\odot$. The models of \citet{Ardestani} strongly depend on the Rossby number, but this is not a critical quantity for deciding whether a planet is tidally locked. We therefore defined a constant Rossby number of 0.5 for unsaturated stars and 0.045 for saturated stars as mean representatives of both regimes. In addition, the models have the following characteristics:

\begin{itemize}
    \item The stellar radius was estimated following \citet{CK17}.
    \item The stellar luminosity was estimated using the $M-L$ empirical relations of \citet{Benedict16} for M-type stars and of \citet{Eker15} for more massive stars.
    \item The boundaries of the HZ were obtained following \citet{Kopparapu13, Kopparapu14} taking into account the "inner edge" displacement caused by the mass of the planet.
    \item To distinguish whether the exoplanet is tidally locked, we used the equations of \citet{Griess09}, assuming an exoplanetary system of 2 Gyr.
    \item The orbital eccentricity was fixed to zero.
    \item For a tidally locked, we assumed a spin-orbit relation of 1:1.
\end{itemize}

With all these assumptions, we placed at the two boundaries of the HZ (D$_2$ and D$_3$) a highly magnetized super-Earth as an example with very high protection estimated at a magnetic polar induction of 3 G when the planet rotates freely.

\begin{sidewaystable*}
	\centering
	\caption{Results obtained for a super-Earth located at the HZ boundaries of simulated stars in the unsaturated regime. See text for details related to the meaning of the different columns.  \label{Tab:no_saturadas_supertierra}
	}
	\begin{tabular}{cccc|cccccc|cccccc}
		\hline
		\multicolumn{4}{c}{Simulated stars} & \multicolumn{5}{c}{Super-Earth at D$_2$} & \multicolumn{5}{c}{Super-Earth at D$_3$}\\
		Mass & Radius & Luminosity & Period & D$_2$ & Tidally & $B_{\rm pp}$ & $r_{\rm M}$/$R_{\rm P}$ & $\alpha_{\rm 0}$ & Prob.  & D$_3$ & Tidally & $B_{\rm pp}$ & $r_{\rm M}$/$R_{\rm P}$ & $\alpha_{\rm 0}$ & Prob.  \\
		(M$_\odot$) & (R$_\odot$) & (L$_\odot$) & (days) & (AU) & locked & (G) & & ($^\circ)$ & & (AU) & locked & (G) & & ($^\circ)$ &\\
		\hline
		\multicolumn{14}{c}{Saturated stars}\\
		\hline
        1 & 1 & 0.94 & 0.6 & 0.895 & N & 3 & 5.9 & 24 & 1 & 1.639 & N& 3 & 7.8 & 21 & 1\\
        0.9 & 0.91 & 0.57 & 0.7 & 0.711& N & 3 & 5.4 & 26 & 1 & 1.321 & N & 3 & 7.1 & 22 & 1\\
        0.8 & 0.82 & 0.32 & 1.0 & 0.547& N & 3 & 4.9 & 27 & 0.97 & 1.033& N & 3 & 6.4 & 23 & 1\\
        0.7 & 0.73 & 0.17 & 1.2 & 0.403& N & 3 & 4.4 & 29 & 0.91 & 0.777& N & 3 & 5.7 & 25 & 1\\
        0.6 & 0.64 & 0.08 & 1.4 & 0.280& Y & 0.07 & 1.1 & 69 & 0 & 0.555& N & 3 & 5.4 & 25 & 1\\
        0.5 & 0.54 & 0.04 & 1.7 & 0.205& Y & 0.09 & 1.3 & 63 & 0 & 0.411& N & 3 & 5.6 & 25 & 1\\
        0.4 & 0.45 & 0.03 & 2.0 & 0.159& Y & 0.11 & 1.3 & 61 & 0 & 0.320& N & 3 & 5.3 & 26 & 1\\
        0.3 & 0.35 & 0.01 & 2.7 & 0.114& Y & 0.14 & 1.3 & 60 & 0 & 0.232& Y & 0.06 & 1.4 & 59 & 0\\
        0.2 & 0.24 & 0.005 & 3.8 & 0.072& Y & 0.21 & 1.4 & 57 & 0 & 0.147& Y & 0.09 & 1.5 & 56 & 0\\
        0.1 & 0.13 & 0.001 & 6.6 & 0.032& Y & 0.41 & 1.6 & 52 & 0 & 0.067& Y & 0.18 & 1.7 & 50 & 0\\
        0.08 & 0.11 & 0.0006 & 7.6 & 0.025& Y & 0.52 & 1.7 & 50 & 0.01 & 0.052& Y & 0.23 & 1.9 & 47 & 0.03\\
        \hline
        \multicolumn{14}{c}{Unsaturated stars}\\
		\hline
        1 & 1 & 0.94 & 6.4 & 0.895 & N & 3 & 9.3 & 19 & 1 & 1.639& N & 3 & 12.1 & 17 & 1\\
        0.9 & 0.91 & 0.57 & 8.1 & 0.711& N  & 3 & 8.5 & 20 & 1 & 1.321& N & 3 & 11.0 & 18 & 1\\
        0.8 & 0.82 & 0.32 & 10.7 & 0.547& N  & 3 & 7.8 & 21 & 1 & 1.033& N & 3 & 10.0 & 18 & 1\\
        0.7 & 0.73 & 0.17 & 13.4 & 0.403& N  & 3 & 7.0 & 22 & 1 & 0.777& N & 3 & 9.0 & 20 & 1\\
        0.6 & 0.64 & 0.08 & 15.5 & 0.280& Y  & 0.07 & 2.0 & 45 & 0.06 & 0.555& N & 3 & 8.4 & 20 & 1\\
        0.5 & 0.54 & 0.04 & 18.4 & 0.205 & Y& 0.09 & 2.2 & 42 & 0.11 & 0.411& N & 3 & 9.0 & 20 & 1\\
        0.4 & 0.45 & 0.03 & 22.7 & 0.159 & Y& 0.11 & 2.2 & 43 & 0.11 & 0.320& N & 3 & 8.6 & 20 & 1\\
        0.3 & 0.35 & 0.013 & 29.6 & 0.114 & Y& 0.14 & 2.2 & 42 & 0.11 & 0.232& Y & 0.06 & 2.1 & 43 & 0.09\\
        0.2 & 0.24 & 0.005 & 42.3 & 0.072 & Y& 0.21 & 2.2 & 42 & 0.11 & 0.147& Y & 0.09 & 2.1 & 43 & 0.09\\
        0.1 & 0.13 & 0.001 & 72.9 & 0.032 & Y& 0.41 & 2.2 & 43 & 0.10 & 0.067& Y & 0.18 & 2.1 & 44 & 0.08\\
        0.08 & 0.11 & 0.0006 & 84.5 & 0.025 & Y& 0.52 & 2.2 & 43 & 0.10 & 0.052& Y & 0.23 & 2.1 & 44 & 0.08\\
		\hline
	\end{tabular}
\end{sidewaystable*}

The results are summarized in Table \ref{Tab:no_saturadas_supertierra} for stars in the saturated and unsaturated regimes. These tables show two very clear regimes in terms of unprotected angle and its translation into an atmosphere-retaining probability. When the planet rotates freely, it is highly magnetized and has an unprotected angle lower than 30$^\circ$ (probability higher than 0.9) in any case, regardless of whether the star is in a saturated regime. As soon as the planet is tidally locked, the unprotected angle increases dramatically to values always higher than 42$^\circ$ with atmosphere-remaining probabilities lower than 0.12.

Using the same set of simulated stars, we also studied the exoplanet magnetic field that is required for a probability of retaining its atmosphere of 0.5 and 1 (i.e., unprotected angles of 34.8$^\circ$ and 25$^\circ$ , respectively) at the two boundaries of the HZ.

\begin{figure*}
 \includegraphics[width=\linewidth]{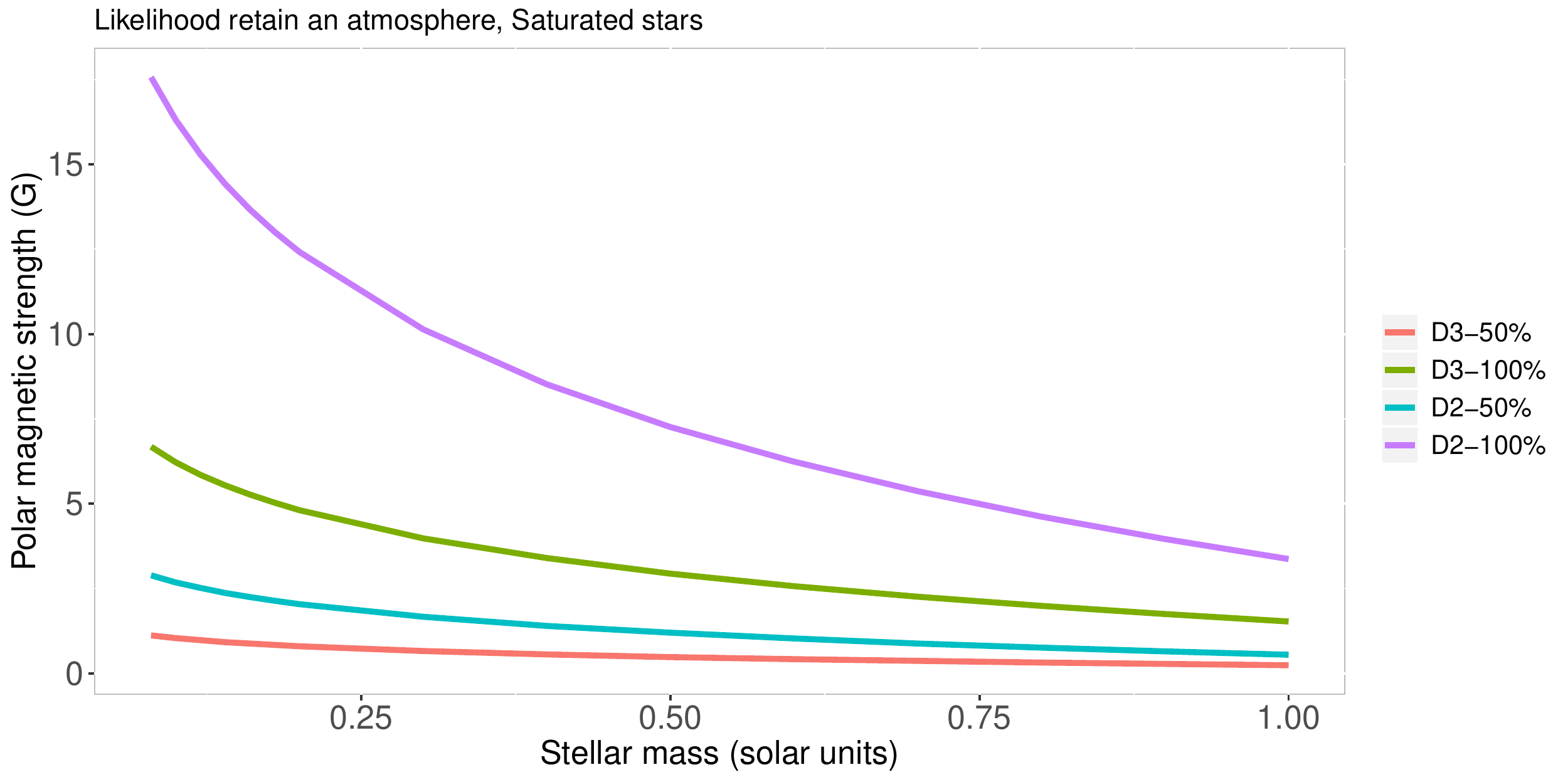}
 \caption{Polar magnetic strength of an exoplanet that is required to ensure a certain probability to retain its atmosphere when it orbits a saturated star at the boundaries of the HZ as a function of stellar mass.}
 \label{Fig:earth}
\end{figure*}

In Fig. \ref{Fig:earth} we show the results obtained at the worst scenario possible, that is, stars in the saturated regime. The polar magnetic strength required for a probability 1 to retain an atmosphere if the planet is located at the inner boundary of the HZ (red line) is really large, especially when the exoplanet orbits a low-mass star. This required strength decreases at the outer boundary of the HZ. A probability of 0.5 to retain an atmosphere requires a much lower polar magnetic strength, with values that can be easily fulfilled by real exoplanets, especially at the outer boundary of the HZ for massive hosting stars.

\subsection{Real cases}

We finally applied our formalism to some real cases. We used the characteristics found at exoplanets.eu data base. We assumed an uncertainty of $\pm 10\%$ for the orbital period of the star when no observational error was given.

We selected a list of 26 known telluric exoplanets orbiting GKM stars with known rotational period. In Table \ref{Tab:real_cases_data} we summarize the main physical characteristics of these exoplanets. Many of them are tidally locked, that is, have synchronous rotation. GJ 3323b and GJ 581e probably rotate in a super-synchronous regime with a 3:2 resonance.

With all these observables and taking into account the orbital eccentricity, we estimated their $\alpha_{\rm 0}$ and the corresponding probability for retaining an atmosphere either at their periastron and apoastron. We found that for a given orbit, $\alpha_{\rm 0}$ may vary by up to 7$^\circ$. These estimates, obtained for the semimajor orbital axis of the planet, are shown in Table \ref{Tab:real_cases_values}.

Analyzing the results, we found that

\begin{itemize}
\item all the planets in free rotation have a probability for retaining an atmosphere close to unity, while tidally locked planets have a very low probability (lower than 0.17). This is a direct consequence of the effect of the planet rotation velocity on the amount of protected atmospheric surface. 
\item exoplanets orbiting saturated stars also have low probabilities of keeping the atmosphere. Because they usually orbit red dwarfs in closed orbits, they are also coupled by tide, and the planets are affected by a combination of effects.
\item all planets that are in orbits close enough to their star to allow the effect of the stellar wind in the sub-Alfv\'enic regime have a very low probability of maintaining their atmosphere.
\end{itemize}

For this final study, we carried out an analysis of the uncertainties of the different variables as a consequence of the observational uncertainties. We assumed a Gaussian distribution of the observational uncertainties, and with a sampling of 25000 random realizations following these probability distributions, we propagated the uncertainties using the equations described in Section 2. This provided a final random distribution of the resulting variables that also followed a Gaussian distribution. The central values displayed at Table \ref{Tab:real_cases_values} are the mean values of these distributions, and the uncertainty is their standard deviation ($\sigma$). The uncertainties coming from the assumptions and simplifications we applied to derive the equations we used can be hardly evaluated, but we assume that the main physics driving the process we described here are contained in our formalism.



\section{Conclusions}

We used different theoretical prescriptions in the literature and proposed a method for estimating the magnetopause radius, the aperture of the auroral ring, and the probability of maintaining the atmosphere of exoplanets orbiting stars with a convective envelope. The magnetopause radius is the mean value of the magnetopause at steady state. Random phenomena such as flares, mass ejections, or other transient events were not taken into account in our analysis.

By means of the dynamical pressure of the stellar wind and the stellar magnetic pressure where the exoplanet is located, we estimated the unprotected atmosphere surface of the exoplanet through the unprotected angle $\alpha_{\rm 0}$ and the probability of the exoplanet to retain its atmosphere compared with the known history of Earth.

This estimation can be made using as input the stellar mass, radius, effective temperature, and rotational period, and on the other hand, the exoplanet mass, radius, orbital distance, and eccentricity, and the following assumptions:

\begin{itemize}
    \item The planet is magentized with a dipolar magnetic field, whose magnetic induction at the poles is known.
    \item The star has an outer convective layer, that is, it has a stellar mass of between 0.08 and 1.3 M$_\odot$ and its axis of rotation is aligned with the magnetic axis of the planet.
    \item The stellar wind is isotropic and isothermal.
\end{itemize}



We analyzed Earth-like planets orbiting the HZ of their hosting star and found that depending on the rotational regime of the star (saturated or unsaturated), the auroral ring and the probability of keeping the atmosphere are clearly divided into two regions above and below 36$^\circ$ and 0.4, respectively. We also found a linear relation between the auroral ring at the inner edge of the HZ ($\alpha_\text{o}(\text{D}_2)$) and the difference auroral ring at the outer and inner edges of the HZ ($\alpha_\text{o}(\text{D}_3)$-$\alpha_\text{o}(\text{D}_2)$). The slope of this linear relation changes with the rotational regime of the star.

We also found that the rotational velocity of the exoplanet is a fundamental factor for the protection of the magnetized planet atmospheres. When the planet is tidally locked, that is, when its rotation velocity is low, the unprotected angle increases dramatically to values higher than 40$^\circ$ with a low likelihood of keeping the atmosphere. In addition, when the effect of the stellar wind on the exoplanet is produced in the sub-Alfv\'enic regime, which commonly occurs in the case of exoplanets orbiting very close to their star, saturated or not, the likelihood of keeping the atmosphere is close to zero.

M-dwarf stars remain in a saturated regime during a significant percentage of their life on the main sequence, which means that the probability that planets like Earth that orbit the HZ of these stars can maintain their atmosphere is reduced. In the particular case of stars with masses lower than 0.3 M$_\odot$, the exoplanets are also probably tidally locked, which dramatically decreases the probability that they can maintain their atmospheres.

We applied our formalism to a set of 26 real telluric exoplanets orbiting GKM stars. This confirmed our previous conclusions: when the star rotates in the saturated regime and/or the planet is tidally locked to its star, the auroral ring has an angle larger than 40$^\circ$ and the probability of retaining its atmosphere is close to zero. We would finally like to remark that the exoplanets with the highest value of the habitability index SEPHI (Kepler-186 f, Kepler 1229 b, and Kepler-442 b) have a probability of maintaining their atmospheres close or equal to one.

\begin{acknowledgements}
AM acknowledges funding from the European Union's Horizon 2020 research and innovation program under the Marie Sklodowska-Curie grant agreement No 749962 (project THOT). The authors also wish to acknowledge the anonymous referee for the constructive comments and suggestions.
\end{acknowledgements}

\begin{appendix}
\section{Additional tables}
\begin{sidewaystable*}
\footnotesize
\begin{threeparttable}
	\caption{Physical characteristics of the stars described in Sect. \ref{sec:ELHZ}. The magnetic fields were obtained through Zeeman-Doppler imaging. The rotation regime consists of the two different regimes described by Eq. \ref{eq:saturation}. $R_{\rm o}$ was calculated with Eqs. \ref{eq:rossby} and \ref{eq:tau}. $D_2$ and $D_3$ are the boundaries of the HZ for an Earth-like exoplanet orbiting the star.} \label{Tab:earth_like_data}
	\centering
	\begin{tabular}{ccccccccccccc}
		\hline
		\multirow{2}{*}{Star} & Spectral & Mass & Radius & Period & $T_{\rm eff}$ & $R_o$ & Magnetic & Rotation & Calculated & $D_2$ & $D_3$\\
		& Type & (in $M_\odot$) & (in $R_\odot$) & (days) & (K) & & Field (G) & Regime & $R_o$ & (AU) & (AU)\\
		\hline
		Sun\tnote{1} & G2V & 1.00 & 1.00 & 28.0 & 5780 & 2.17 & - & Unsaturated & 2.17 & 0.95 & 1.68\\
		GJ 182\tnote{2} & M0.5 & 0.75 & 0.82 & 4.4 & 3950 & 0.17 & 172 & Unsaturated & 0.17 & 0.40 & 0.75\\
		DT Vir\tnote{2} & M0.5 & 0.59 & 0.53 & 2.9 & 3790 & 0.09 & 149 & Saturated & 0.09 & 0.24 & 0.45\\
		DS Leo\tnote{2} & M0 & 0.58 & 0.52 & 15.7\tnote{7} & 3770 & - & 87 & Unsaturated & 0.50 & 0.23 & 0.44\\
		GJ 49\tnote{2} & M1.5 & 0.57 & 0.51 & 0.74\tnote{8} & 3750 & - & 27 & Saturated & 0.02 & 0.22 & 0.43\\
		OT Ser\tnote{2} & M1.5 & 0.55 & 0.49 & 3.4 & 3690 & 0.10 & 123 & Saturated & 0.10 & 0.21 & 0.40\\
		CE Boo\tnote{2} & M2.5 & 0.48 & 0.43 & 14.7 & 3570 & 0.35 & 103 & Unsaturated & 0.39 & 0.17 & 0.33\\
		AD Leo\tnote{3} & M3 & 0.42 & 0.38 & 2.24 & 3540 & 0.05 & 180 & Saturated & 0.05 & 0.15 & 0.29\\
		EQ Peg A\tnote{3} & M3.5 & 0.39 & 0.35 & 1.06 & 3530 & 0.02 & 480 & Saturated & 0.02 & 0.14 & 0.26\\
		EV Lac\tnote{3} & M3.5 & 0.32 & 0.30 & 4.4 & 3480 & 0.07 & 490 & Saturated & 0.08 & 0.11 & 0.22\\
		V374 Peg\tnote{3} & M4 & 0.28 & 0.28 & 0.45 & 3410 & 0.006 & 640 & Saturated & 0.007 & 0.10 & 0.20\\
		EQ Peg B\tnote{3} & M4.5 & 0.25 & 0.25 & 0.4 & 3340 & 0.005 & 450 & Saturated & 0.006 & 0.09 & 0.17\\
		GJ 1156\tnote{4} & M5 & 0.14 & 0.16 & 0.49 & 3160 & 0.005 & 100 & Saturated & 0.004 & 0.05 & 0.10\\
		GJ 1245 B\tnote{4} & M5.5 & 0.12 & 0.14 & 6.87\tnote{8} & 3030 & - & 60 & Saturated & 0.006 & 0.04 & 0.08\\
		WX Uma\tnote{4} & M6 & 0.1 & 0.12 & 0.78 & 2770 & 0.008 & 1060 & Saturated & 0.006 & 0.03 & 0.06\\
		DX Cnc\tnote{4} & M6 & 0.1 & 0.11 & 0.46 & 2770 & 0.005 & 80 & Saturated & 0.003 & 0.03 & 0.05\\
		Kepler-22\tnote{9} & G5 & 0.97 & 0.98 & 25.6 & 5518 & - & - & Unsaturated & 2.08 & 0.86 & 1.53\\
		Kepler-1185\tnote{9} & - & 0.96 & 0.87 & 12.7 & 5622 & - & - & Unsaturated & 0.78 & 0.79 & 1.40\\
		Kepler-62\tnote{9} & K2V & 0.69 & 0.63 & 36.0 & 4869 & - & - & Unsaturated & 1.34 & 0.45 & 0.81\\
		K2-3\tnote{9} & M0V & 0.61 & 0.55 & 10.1 & 3951 & - & - & Unsaturated & 0.34 & 0.27 & 0.51\\
		Kepler-442\tnote{9} & - & 0.61 & 0.60 & 48.7 & 4402 & - & - & Unsaturated & 1.60 & 0.35 & 0.66\\
		Kepler-1229\tnote{9} & - & 0.54 & 0.51 & 17.7 & 3784 & - & - & Unsaturated & 0.52 & 0.23 & 0.43\\
		Kepler-186\tnote{9} & M1V & 0.48 & 0.47 & 34.3 & 3788 & - & - & Unsaturated & 0.90 & 0.21 & 0.40\\
		GJ 581\tnote{9} & M2.5V & 0.31 & 0.30 & 132.5 & 3498 & - & - & Unsaturated & 2.36 & 0.11 & 0.22\\
		Wolf 1061\tnote{9} & M3V & 0.25 & 0.31 & 119.3 & 3342 & - & - & Unsaturated & 1.71 & 0.11 & 0.21\\
		Kepler-1649\tnote{9} & M5V & 0.22 & 0.25 & 57.5 & 3240 & - & - & Unsaturated & 0.74 & 0.08 & 0.16\\
		Ross 128\tnote{9} & M4V & 0.17 & 0.20 & 0.5 & 3192 & - & - & Saturated & 0.01 & 0.06 & 0.12\\
		GJ 3323\tnote{9} & M4V & 0.16 & 0.12 & 88.5 & 3159 & - & - & Unsaturated & 1.03 & 0.04 & 0.07\\
		Barnard's\tnote{10} & M3.5V & 0.16 & 0.18 & 140.0 & 3278 & - & - & Unsaturated & 1.44 & 0.06 & 0.12\\
		YZ Cet\tnote{9} & M4.5V & 0.13 & 0.17 & 69.2 & 3056 & - & - & Unsaturated & 0.57 & 0.05 & 0.10\\
		Prox. Cen\tnote{5} & M5.5 & 0.12 & 0.14 & 83.0 & 3050 & 0.65 & - & Unsaturated & 0.67 & 0.04 & 0.08\\
        Teegarden's\tnote{11} & M7.0V & 0.089 & 0.107 & 100 & 2904 & - & - & Unsaturated & 0.66 & 0.03 & 0.06\\
        Trappist-1\tnote{6} & M8 & 0.08 & 0.117 & 3.3 & 2550 & 0.03 & - & Saturated & 0.02 & 0.02 & 0.05\\
		\hline
	\end{tabular}
	\begin{tablenotes}
\item[1] \citet{Matt15} 
\item[2] \citet{Donati08} 
\item[3] \citet{Morin08}
\item[4] \citet{Morin10}
\item[5] \citet{Anglada16}
\item[6] \citet{Gillon17}
\item[7] \citet{Reiners18}
\item[8] \citet{Newton16}
\item[9] http://exoplanet.eu/
\item[10] \citet{Ribas18}
\item[11] \citet{Zechmeister19}
\end{tablenotes}
\end{threeparttable}
\end{sidewaystable*}

\begin{table}
	\centering
	\caption{Estimated values for the stars of Table \ref{Tab:earth_like_data} and an exoplanet at the inner boundary (D$_2$) and the outer boundary (D$_3$) of the HZ. $r_{\rm M}$/$R_{\rm P}$ is the radius of the exoplanetary magnetopause relative to the exoplanetary radius. $\alpha_{\rm 0}$ is the aperture angle of the auroral ring, "Prob." is the likelihood of the exoplanet to retain its atmosphere.	\label{Tab:earth_like}
	}
	\begin{tabular}{c|ccc|ccc}
		\hline
		& \multicolumn{3}{c}{Results at D$_2$} & \multicolumn{3}{c}{Results at D$_3$}\\
		\hline
		\multirow{2}{*}{Star} & $r_{\rm M}$/$R_{\rm P}$ & $\alpha_{\rm 0}$ & Prob.  & $r_{\rm M}$/$R_{\rm P}$ & $\alpha_{\rm 0}$ & Prob.\\
		&  & (in $^\circ$) & &  & (in $^\circ$) & \\
		\hline
		Sun & 9.8 & 19 & 1.0 & 12.1 & 16 & 1.0\\
		GJ 182  & 3.6 & 32 & 0.70 & 4.6 & 28 & 0.95\\
		DT Vir  & 2.7 & 38 & 0.32 & 3.4 & 33 & 0.66\\
		DS Leo  & 4.3 & 29 & 0.91 & 5.5 & 25 & 1.0 \\
		GJ 49  & 2.7 & 38 & 0.32 & 3.5 & 33 & 0.67\\
		OT Ser  & 2.7 & 38 & 0.33 & 3.5 & 32 & 0.68\\
		CE Boo  & 4.6 & 28 & 0.95 & 5.9 & 24 & 1.0\\
		AD Leo  & 2.8 & 37 & 0.36 & 3.7 & 31 & 0.75\\
		EQ Peg A  & 2.8 & 37 & 0.36 & 3.7 & 32 & 0.74\\
		EV Lac & 2.7 & 37 & 0.33 & 3.6 & 32 & 0.71\\
		V374 Peg  & 2.6 & 38 & 0.30 & 3.5 & 32 & 0.68\\
		EQ Peg B  & 2.6 & 39 & 0.27 & 3.4 & 33 & 0.65\\
		GJ 1156  & 2.4 & 40 & 0.20 & 3.2 & 34 & 0.57\\
		GJ 1245 B  & 2.3 & 41 & 0.16 & 3.2 & 34 & 0.54\\
		WX Uma  & 2.2 & 43 & 0.10 & 3.0 & 35 & 0.46\\
		DX Cnc  & 2.1 & 44 & 0.09 & 3.0 & 35 & 0.46\\
		Kepler-22  & 9.1 & 19 & 1.0 & 11.6 & 17 & 1.0\\
		Kepler-1185  & 7.0 & 22 & 1.0 & 8.9 & 20 & 1.0\\
		Kepler-62  & 6.6 & 23 & 1.0 & 8.3 & 20 & 1.0\\
		K2-3  & 4.0 & 30 & 0.83 & 5.0 & 27 & 0.98\\
		Kepler-442  & 6.4 & 23 & 1.0 & 8.0 & 21 & 1.0\\
		Kepler-1229  & 4.7 & 28 & 0.95 & 5.9 & 24 & 1.0\\
		Kepler-186  & 5.3 & 26 & 1.0 & 6.7 & 23 & 1.0\\
		GJ 581  & 4.8 & 27 & 0.96 & 5.9 & 24 & 1.0\\
		Wolf 1061  & 4.6 & 28 & 0.94 & 5.7 & 25 & 1.0\\
		Kepler-1649  & 4.0 & 30 & 0.83 & 5.0 & 27 & 0.98\\
		Ross 128  & 2.4 & 40 & 0.21 & 3.3 & 34 & 0.59\\
		GJ 3323  & 3.2 & 34 & 0.58 & 4.0 & 30 & 0.84\\
		Barnard's  & 3.8 & 31 & 0.78 & 4.7 & 27 & 0.96\\
		YZ Cet  & 3.4 & 33 & 0.63 & 4.3 & 29 & 0.89\\
		Prox. Cen  & 3.2 & 34 & 0.57 & 4.0 & 30 & 0.84\\
		Teegarden's & 2.9 & 36 & 0.41 & 3.6 & 32 & 0.72\\
		Trappist-1  & 2.0 & 45 & 0.06 & 2.8 & 37 & 0.39\\
		\hline
	\end{tabular}
\end{table}

\begin{sidewaystable*}
\footnotesize
\begin{threeparttable}
	\caption{Physical characteristics of the exoplanets we used in the study of some real cases. When planetary mass and radius were unknown, we estimated them based on results of \citet{CK17}. The rotational period of tidally locked exoplanets was obtained using the most likely spin-orbit resonance according to \citet{Dobrovolskis07}. The rest of the columns are described in the text.} \label{Tab:real_cases_data}
	\centering
	\begin{tabular}{ccccccccc}
		\hline
		\multirow{2}{*}{Exoplanet} & Mass & Radius & Semi-major & HZ & Mean Orbital & Rotation & Tidally & Magnetic\\
		& (in $M_\oplus$) & (in $R_\oplus$) & axis (AU) & & Eccentricity & (days) & Locked & Moment\\
		\hline
		Barnard's b\tnote{a} & 3.23 & 1.60\tnote{e} & 0.404 & COLD & 0.295 & - & N & 2.47\\
		GJ 2232 b\tnote{b} & 2.02 & 1.22\tnote{e} & 0.033 & INNER EDGE & 0.230 & 3.576 & Y & 0.70\\
		GJ 2232 c\tnote{b} & 2.31 & 1.31\tnote{e} & 0.126 & COLD & 0.215 & 40.54 & Y & 0.04\\
		GJ 581 e\tnote{b} & 1.94 & 1.20\tnote{e} & 0.028 & HOT & 0.320 & 2.100 & Y & 1.02\\
		K2-3 d\tnote{b} & 11.09 & 1.48 & 0.209 & INNER EDGE& 0.045 & 44.56 & Y & 2.2\\
		Kepler-1185 b\tnote{b} & 2.29\tnote{e} & 1.31 & 0.428 & INNER EDGE & 0 & - & N & 0.76\\
		Kepler-1229 b\tnote{b} & 2.49\tnote{e} & 1.37 & 0.311 & GREEN & 0 & - & N & 0.19\\
		Kepler-1649 b\tnote{b} & 1.22\tnote{e} & 1.06 & 0.050 & HOT & 0 & 8.689 & Y & 0.22\\
		Kepler-186 f\tnote{b} & 1.34\tnote{e} & 1.09 & 0.392 & GRENN & 0 & - & N & 1.3\\
		Kepler-22 b\tnote{b} & 6.00\tnote{e} & 2.30 & 0.848 & GREEN & 0 & - & N & 0.77\\
		Kepler-442 b\tnote{b} & 2.33\tnote{e} & 1.32 & 0.386 & GREEN & 0.04 & - & N & 8.4\\
		Kepler-62 e\tnote{b} & 35.91\tnote{e} & 1.58 & 0.426 & INNER EDGE & 0 & - & N & 92.0\\
		Kepler-62 f\tnote{b} & 35.96\tnote{e} & 1.38 & 0.718 & GREEN & 0 & - & N & 68.0\\
		Proxima Cen. b\tnote{c} & 1.27 & 1.07\tnote{e} & 0.048 & GREEN & 0.175 & 11.186 & Y & 0.19\\
		Ross 128 b\tnote{b} & 1.40 & 1.10\tnote{e} & 0.050 & HOT & 0.116 & 9.866 & Y & 0.23\\
		Teegarden's b\tnote{b} & 1.05 & 1.01\tnote{e} & 0.025 & INNER EDGE & 0.08 & 4.910 & Y & 0.30\\
		Teegarden's c\tnote{b} & 1.11 & 1.03\tnote{e} & 0.044 & GREEN & 0.08 & 11.409 & Y & 0.16\\
		Trappist-1 c\tnote{d} & 1.38 & 1.06 & 0.015 & HOT & 0.042 & 2.422 & Y & 1.10\\
		Trappist-1 d\tnote{d} & 0.41 & 0.77 & 0.021 & HOT & 0.035 & 4.050 & Y & 0.15\\
		Trappist-1 e\tnote{d} & 0.64 & 0.90 & 0.028 & GREEN & 0.043 & 6.100 & Y & 0.10\\
		Trappist-1 g\tnote{d} & 1.34 & 1.13 & 0.045 & GREEN & 0.031 & 12.353 & Y & 0.06\\
		Trappist-1 h\tnote{d} & 0.37\tnote{e} & 0.76 & 0.066 & OUTER EDGE & 0.0 & 20.000 & Y & 0.03\\
		Wolf 1061 b\tnote{b} & 1.91 & 1.20\tnote{e} & 0.036 & HOT & 0.165 & 4.887 & Y & 0.52\\
		YZ Cet b\tnote{b} & 0.76 & 0.93\tnote{e} & 0.016 & HOT & 0.060 & 1.969 & Y & 0.46\\
		YZ Cet c\tnote{b} & 0.99 & 1.0\tnote{e} & 0.021 & HOT & 0.055 & 3.060 & Y & 0.46\\
		YZ Cet d\tnote{b} & 1.14 & 1.04\tnote{e} & 0.028 & HOT & 0.129 & 4.656 & Y & 0.33\\
		\hline
	\end{tabular}
	\begin{tablenotes}
\item[a] \citet{Ribas18} 
\item[b] http://exoplanet.eu/catalog/ 
\item[c] \citet{Anglada16}
\item[d] \citet{Gillon17}
\item[e] Esitmated using \citet{CK17}
\end{tablenotes}
\end{threeparttable}
\end{sidewaystable*}

\begin{table*}
	\caption{Results obtained using our formalism for various telluric exoplanets. $\alpha_{\rm 0, peri}$ - $\alpha_{\rm 0, apo}$ is the difference in aperture angles between the orbital periastron and apastron. The rest of the columns are described in the text.} \label{Tab:real_cases_values}
	\centering
	\begin{tabular}{c|ccccc}
		\hline
		\multirow{2}{*}{Exoplanet} & Wind & $r_{\rm M}$/$R_{\rm P}$ & $\alpha_{\rm 0}$ & Prob. & $\alpha_{\rm 0, peri}$ - $\alpha_{\rm 0, apo}$\\
		& Impact &  & ($^\circ$) &  & ($^\circ$)\\
		\hline
		Barnard's b & Super-Alf. & 7.91$\pm$0.16 & 20.84$\pm$0.22 & 1.0$\pm$0.0 & 2.4$\pm$0.6\\
		GJ 3323 b & Super-Alf. & 2.39$\pm$0.05 & 40.3$\pm$0.5 & 0.19$\pm$0.02 & 3.6$\pm$1.0\\
		GJ 3323 c & Super-Alf. & 1.88$\pm$0.04 & 46.8$\pm$0.6 & 0.03$\pm$0.01 & 4.5$\pm$2.1\\
		GJ 581 e & Super-Alf. & 2.69$\pm$0.06 & 37.6$\pm$0.5 & 0.32$\pm$0.03 & 4.8$\pm$0.9\\
		K2-3 d& Super-Alf. & 1.83$\pm$0.08 & 47.8$\pm$1.4 & 0.03$\pm$0.01 & 1.1$\pm$0.6\\
		Kepler-1185 b& Super-Alf. & 5.63$\pm$0.15 & 24.9$\pm$0.3 & 1.0$\pm$0.0 & 0.0$\pm$0.0\\
		Kepler-1229 b& Super-Alf. & 4.98$\pm$0.23 & 26.6$\pm$0.7 & 0.98$\pm$0.01 & 0.0$\pm$0.0\\
		Kepler-1649 b& Super-Alf. & 1.95$\pm$0.05 & 45.8$\pm$0.7 & 0.05$\pm$0.01 & 0.0$\pm$0.0\\
		Kepler-186 f& Super-Alf. & 6.80$\pm$0.15 & 22.6$\pm$0.3 & 1.0$\pm$0.0 & 0.0$\pm$0.0\\
		Kepler-22 b& Super-Alf. & 9.3$\pm$0.3 & 19.1$\pm$0.3 & 1.0$\pm$0.0 & 0.0$\pm$0.0\\
		Kepler-442 b& Super-Alf. & 11.9$\pm$0.3 & 16.83$\pm$0.20 & 1.0$\pm$0.0 & 0.24$\pm$0.15\\
		Kepler-62 e& Super-Alf. & 10.2$\pm$0.3 & 18.3$\pm$0.2 & 1.0$\pm$0.0 & 0.0$\pm$0.0\\
		Kepler-62 f& Super-Alf. & 12.7$\pm$0.3 & 16.32$\pm$0.21 & 1.0$\pm$0.0 & 0.0$\pm$0.0\\
		Proxima Cen b& Super-Alf. & 1.86$\pm$0.04 & 47.3$\pm$0.7 & 0.03$\pm$0.01 & 3.7$\pm$2.2\\
		Ross 128 b& Sub-Alf. & 1.25$\pm$0.03 & 63.6$\pm$1.3 & 0.0$\pm$0.0 & 7$\pm$4\\
		Teegarden's b& Super-Alf. & 1.87$\pm$0.04 & 47.1$\pm$0.7 & 0.03$\pm$0.01 & 1.6$\pm$0.9\\
		Teegarden's c& Super-Alf. & 1.81$\pm$0.04 & 48.1$\pm$0.6 & 0.02$\pm$0.0 & 1.7$\pm$1.0\\
		Trappist-1 c& Sub-Alf. & 1.26$\pm$0.03 & 62.8$\pm$1.3 & 0.0$\pm$0.0 & 2.9$\pm$1.7\\
		Trappist-1 d& Sub-Alf. & 1.19$\pm$0.03 & 66.4$\pm$1.5 & 0.0$\pm$0.0 & 2.7$\pm$1.6\\
		Trappist-1 e& Sub-Alf. & 1.27$\pm$0.03 & 62.8$\pm$1.3 & 0.0$\pm$0.0 & 2.6$\pm$1.5\\
		Trappist-1 g& Super-Alf. & 1.38$\pm$0.03 & 58.3$\pm$1.1 & 0.0$\pm$0.0 & 1.3$\pm$0.7\\
		Trappist-1 h& Super-Alf. & 1.35$\pm$0.15 & 61$\pm$7 & 0.0$\pm$0.0 & 0.0$\pm$0.0\\
		Wolf 1061 b& Super-Alf. & 2.25$\pm$0.05 & 41.9$\pm$0.5 & 0.01$\pm$0.02 & 2.9$\pm$1.2\\
		YZ Cet b& Sub-Alf. & 1.84$\pm$0.05 & 47.6$\pm$0.8 & 0.03$\pm$0.01 & 1.7$\pm$1.0\\
		YZ Cet c& Sub-Alf. & 1.88$\pm$0.04 & 46.9$\pm$0.7 & 0.03$\pm$0.01 & 1.8$\pm$1.1\\
		YZ Cet d& Super-Alf. & 1.89$\pm$0.04 & 46.7$\pm$0.7 & 0.03$\pm$0.02 & 2.9$\pm$1.3\\
		\hline
	\end{tabular}
\end{table*}

\end{appendix}

\end{document}